\documentclass[]{article}
\usepackage{graphicx}

\begin{document}

\centerline{\Large \bf Fragmentation Experiment and Model}
\centerline{\Large \bf for Falling Mercury Drops}

\bigskip

\centerline{P.M.C. de Oliveira$^1$, C.A.F. Leite$^1$, C.V. Chianca$^1$,
J.S. S\'a Martins$^1$}
\centerline{and C.F. Moukarzel$^2$}

\bigskip

$^1$ Instituto de F\'\i sica, Universidade Federal Fluminense\par

av. Litor\^anea s/n, Boa Viagem, Niter\'oi, Brasil 24210-340

\medskip

$^2$ Departamento de Fisica Aplicada, CINVESTAV Unidad Merida\par

Antigua Carretera a Progresso, km 6, 97310 Merida, Mexico

\bigskip
e-mail address: pmco@if.uff.br

\bigskip

\begin{abstract}

	The experiment consists of counting and measuring the size of the many
fragments observed after the fall of a mercury drop on the floor. The size
distribution follows a power-law for large enough fragments. We address the
question of a possible crossover to a second, different power-law for small
enough fragments. Two series of experiments were performed. The first uses a
traditional film photographic camera, and the picture is later treated on a
computer in order to count the fragments and classify them according to their
sizes. The second uses a modern digital camera. The first approach has the
advantage of a better resolution for small fragment sizes. The second, although
with a poorer size resolution, is more reliable concerning the counting of {\sl
all} fragments up to its resolution limit. Both together clearly indicate the
real existence of the quoted crossover.

	The model treats the system microscopically during the tiny time
interval when the initial drop collides with the floor. The drop is modelled by
a connected cluster of Ising spins pointing {\sl up} (mercury) surrounded by
Ising spins pointing {\sl down} (air). The Ising coupling which tends to keep
the spins segregated represents the surface tension. Initially the cluster
carries an extra energy equally shared among all its spins, corresponding to
the coherent kinetic energy due to the fall. Each spin which touches the floor
loses its extra energy transformed into a thermal, incoherent energy
represented by a temperature used then to follow the dynamics through Monte
Carlo simulations. Whenever a small piece becomes disconnected from the big
cluster, it is considered a fragment, and counted. The results also indicate
the existence of the quoted crossover in the fragment-size distribution.

\end{abstract}

\newpage

	In many instances, the fracture of solid objects or dense liquids leads
to scaleless distributions of fragment sizes (see, for instance,
\cite{Oddershede}). For a recent study, as well as a nice theoretical analysis,
see also \cite{Astrom}. This scaling phenomenon is observed among other
experiments after the drop of objects on the floor. Objects with different
shapes (cubes, rods, discs, plates, etc) and made of several materials (various
plastics, frozen potatoes, soap, etc) were studied. The main result of each
experiment is the distribution of fragment sizes, classified according to their
masses, which follows a power-law

$$ \frac{N}{m}\, \propto\, m^{-\beta}\,\,\,\, ,$$

\noindent where $N(m)$ is the number of fragments with masses larger than $m$.
Plotted on a double-logarithmic scale, such a distribution shows a decreasing
straight line, the slope of which gives the characteristic exponent $\beta$ of
that particular experiment, $\beta > 1$. This exponent, however, depends on
various parameters, such as the geometry of the initial object, the material
from which it is made, the falling height and others. It is not a universal
index. For the largest fragment sizes, of course, an exponential decay bends
the straight line downwards, but this finite size effect is out of the scope of
the present work.

	The same phenomenon is observed in many other systems within completely
different size scales, including nuclear multifragmentation (see \cite{Campi}
and references therein), atomic cluster fragmentation \cite{Kuhle}, magnetic
liquids \cite{Oddershede2}, icebergs \cite{Korsnes} and meteorite showers
\cite{Oddershede3}.

	In many cases, instead of a single straight line denoting a simple
power-law, one observes two straight lines with different slopes. There is a
crossover from the large-fragment regime (exponent $\beta$ already mentioned)
to the small-fragment counterpart (with a smaller exponent $\bar{\beta}$). This
is the case, among others, of some meteorite showers, fitted in
\cite{Oddershede3} by the sum of two power-laws.

	We raise here the hypothesis that this behaviour, with two superimposed
power-laws, can also be applied to the beautiful experiment carried-out by
Sotolongo-Costa {\sl et al} \cite{Sotolongo} with the fall of mercury drops.
They count and measure the fragment sizes on a microscope, find the expected
power-law behaviour for large fragments, but their resolution is not enough to
sample the supposed small-fragment regime. In order to probe this hypothesis,
we repeat the same experiment with a larger mercury drop (about $0.7$cm
diameter) and observe the result by taking pictures, instead of using a
microscope. The results are shown in figure 1. Along the vertical axis we plot
directly the (normalised) counting $N(m)$ (instead of $N/m$), thus the slopes
correspond to $1-\beta$ instead of $-\beta$.


	Figure 1 shows two different experiments, the first corresponding to
the open circles. Pictures were taken with a traditional camera, using films
ORWO with 50 ASA. This fine grain film has a good resolution, allowing us to
probe a much larger range of mercury fragments, their masses covering 4 orders
of magnitude. (We adopt the smallest mass as unit, in Figure 1). The expected
two regimes appear with the second, less-slanted straight line on the left
describing the smaller fragments. The crossover point falls near mass $m =
10^2$, two orders of magnitude smaller than the largest fragment. The
resolution, however, is enough to probe also fragments two further orders of
magnitute smaller. The plot corresponds to 10 realisations of the experiment,
for an initial drop height of 2.2m, all 10 data sets superimposed. Other
realizations with a height of 2.4m confirm the same behaviour, again with the
film camera. This two-slopes behaviour was also verified with other heights,
with a digital camera, as shown later. Based on these data (circles in figure
1) one would be tempted to conclude in favour of the real existence of the
crossover. Prudence is better. Although presenting good resolution, the film
pictures could in principle suffer from a drawback, as follows. The light
coming from each mercury fragment impresses a cluster of neighbouring grains on
the film surface. The smallest mercury fragments correspond to the smallest
clusters of grains. With a small number of grains to be impressed by the light,
the failure probability of the whole cluster maybe large enough to artificially
increase the undercounting on the region of the smallest fragments. Remember
that the chemical blow-up of a single grain is not free of correlations
concerning the blow-up of neighbouring grains on the film.

\begin{figure}

\begin{center}
 \includegraphics[angle=-90,scale=0.5]{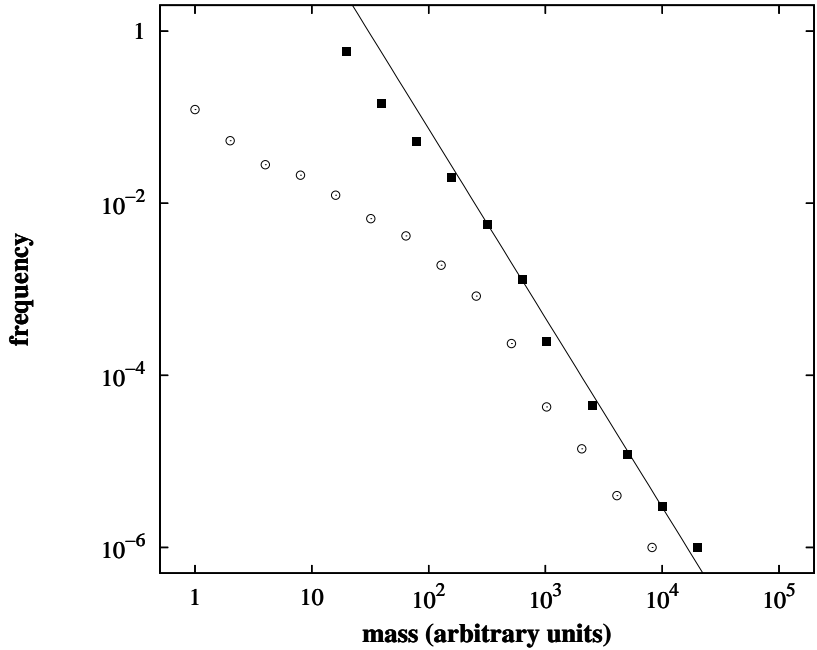}
\end{center}

\caption{Experimental results for the falling mercury drop. Circles correspond
to a traditional camera with good resolution film. Squares were obtained with a
digital camera with greater statistics (a 10-fold larger number of realisations
of the fall experiment). Thus, the largest fragment observed among all digital
pictures is largest than that on the photographic films.}

\label{fig1}
\end{figure}


	Because of this possible drawback, we decided to repeat the experiment
once more, now taking pictures with a digital camera. Its resolution is 3.2
Megapixels. Different heights present always two slopes. As an example, the
results for the same height of 2.2m are shown by the full squares in the same
figure 1. Now, the data correspond to 100 realisations, thus the largest
fragment is larger than the previous one pictured by the traditional camera
with only 10 realisations. The mass ratio between these two largest fragments
is used in order to properly plot both data sets on the same figure with the
same mass unit. The continuous line is a fit for masses larger than $m = 10^2$.
The digital image has a lower resolution than the photographic film, and then
misses fragments smaller than $m = 10$. However, neighbouring pixels are free
of correlations with other pixels because they are independently treated by the
electronic mechanism. Therefore, the probability of undercounting the smaller
mercury fragments is negligible, and we can trust the data in between $m = 10$
and $m = 10^2$. Although small, the bend is also visible. We can safely
conclude that the observation of two distinct power-law exponents is not an
artifact of the measurement process. Note also that both experiments show the
same position for the crossover point, near $m = 10^2$.


	Possibly the same qualitative behaviour would appear with liquids other
than mercury, with different densities and surface tension properties.
Following \cite{Sotolongo}, we used mercury because the relatively high surface
tension keeps the fragments spherical, allowing the experimental determination
of their masses from the two-dimensional projection, i.e. the pictures.
Another, more complicated experimental approach would be necessary for other
liquids. However, within our simulational model described below, one can vary 
the (equivalent of the) surface tension and height at will.

	The collective behaviour of the many microscopic components of the
initial mercury drop, during the tiny time interval when the crash occurs,
should be responsible for the observed fragment distribution. The kinetic
energy aquired during the fall is distributed among these many components,
partially breaking the surface tension which glues them together in a single
drop before the crash. The consequence is to separate it into many different
fragments.

	We decide to treat this behaviour through a simple dynamic lattice drop
model already adopted in many other instances (for a review, see \cite{drop}).
The model consists in treating a drop, with its characteristic surface tension,
as a cluster of neighbouring Ising spins pointing {\sl up} surrounded by spins
pointing {\sl down}. These spins occupy a square or cubic lattice, and do not
move. They can only flip their {\sl up} or {\sl down} orientations. In most
cases the total drop mass should be conserved, then any {\sl up} $\to$ {\sl
down}$\,$ flip is necessarily followed by a {\sl down} $\to$ {\sl up}$\,$ flip
performed on another lattice site. The Ising interaction applies only to
neighbouring spin pairs: if they are parallel, no energy contribution from this
pair is counted; otherwise, if they point one {\sl up} and the other {\sl
down}, then the pair contributes with an energy unit to the whole system. Thus,
the Ising energy is distributed exclusively along the drop surface, i.e. the
interface between the cluster of spins {\sl up} and the sea of spins {\sl down}
outside, and characterises the surface tension. Because of that feature, during
the Monte Carlo dynamics, we toss to flip only pairs of non-neighbouring spins
which are currently positioned along this interface, one pointing {\sl up} and
the other {\sl down}. The interaction neighbourhood we adopt for each spin
corresponds to its 8 (18) nearest neighbours on the square (cubic) lattice.

	This lattice drop model was introduced \cite{Manna} within a static
version in order to study the shape of water drops hanging from a vertical
wall. Later, it was generalised \cite{faucet} to study a dynamic phenomenon:
the time interval series of the successive drops falling from a leaky faucet.
It was also applied to other dynamic phenomena such as nuclear
multifragmentation \cite{nuclear}, magnetic hysteresis \cite{hysteresis} and
annealing phenomena in magnetic multilayers \cite{multilayers}. In each case,
further particular ingredients describing the physical system are introduced,
besides the surface tension common to all cases, represented by the Ising
interaction. Here, we will describe only the present case of the mercury drop.


	We start the computer simulation at the very moment when the initial
mercury ball touches the floor. It is represented by a big round cluster of $N$
spins {\sl up}, and the floor by an inert flat boundary. Our first parameter
$E$ corresponds to the initial kinetic energy accumulated during the fall,
therefore this parameter $E$ represents the falling height. We assign the share
$e = E/N$ to each pixel of the drop. The initial Monte Carlo temperature is $T
= 0$, because no random motion exists at this moment.

	From this starting configuration, we simulate the collision with the 
floor as follows.

\begin{enumerate}

\item The $n$ pixels currently touching the floor are transferred to random
positions along the drop free surface, and the drop as a whole is moved down
one lattice row;

\item The energy $e$ carried by each of these $n$ pixels is set to zero,
because they just lost their coherent initial kinetic energy;

\item The same energy total amount $\Delta{E} = ne$ is transformed into
incoherent, thermal form, by increasing the Monte Carlo temperature $T \to T +
\Delta{E}$;

\item The drop shape is then allowed to relax, by performing $r$ movements on
its free surface, where the number $r$ is our second parameter defining the
relaxation time scale. Each movement consists in flipping two non-neighbouring
spins, one pointing {\sl up} and positioned along the inner free surface, the
other pointing {\sl down} and located along the outer surface. The positions of
both are randomly chosen, and the double flipping is performed or not according
to Metropolis algorithm: if the total Ising energy decreases as a consequence
of this movement, then it is performed with certainty; otherwise, if the total
Ising energy would increase by an amoung $\delta{E}$, then the movement is
performed only with probability $\exp(-\delta{E}/T)$.

\end{enumerate}


	These four steps are repeated iteratively. At some moment during this
process, some piece of the drop becomes disconnected from the main part still
touching the floor. It is considered a fragment. If the total energy of the
whole system increases due to this fragmentation, then the fragment is
discarded and counted in the statistics. Also in this case, the Monte Carlo
temperature is decreased $T \to T - kE_f$, where $E_f$ is the surface Ising
energy carried out by the fragment, and $k$ is our third constant parameter. It
can be varied in order to simulate different surface tensions. The series of
$r$ relaxations proceeds with the remainder drop. The process stops when the
drop vanishes, or when the Monte Carlo temperature is low enough to forbid any
further fragmentation.

	The whole dynamics is repeated again and again, always starting from
the same initial drop with the same initial kinetic energy $E$, and the
fragments are classified according to their masses. We have performed a lot of
tests with different values for our parameters $E$, $r$ and $k$, for both the
square and cubic lattices. In all cases we obtained the same kind of fragment
distribution, showing the already quoted crossover from large to small size
regimes. Figures 2 and 3 are typical examples. The bend is always present. The
two slopes, however, depend slightly on the parameters, characterising their
non-universality generally observed in many other fragmentation problems. Of
course, we have a much better statistics than the experimental set up, thus
these plots present a much greater number of points without need of dividing
the mass axis in bins.

\begin{figure}

\begin{center}
 \includegraphics[angle=-90,scale=0.5]{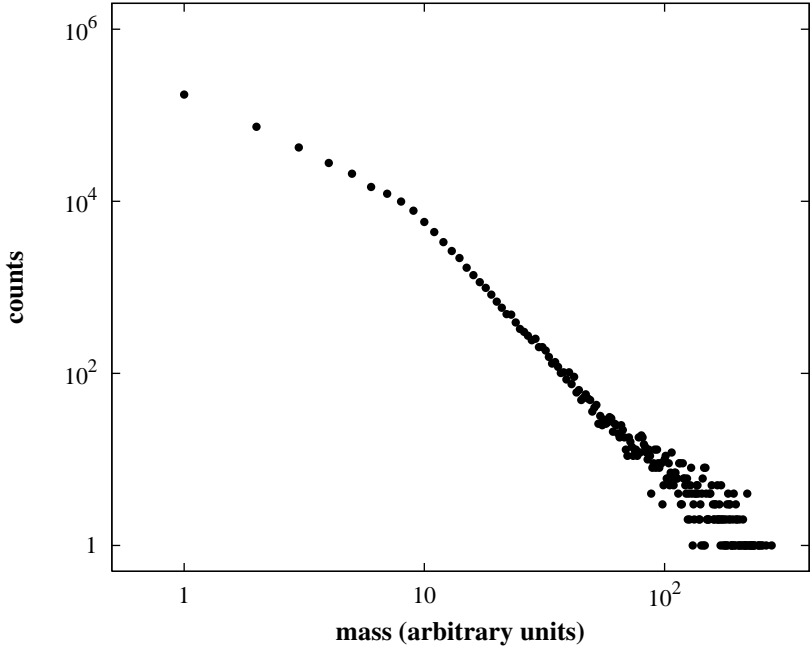}
\end{center}

\caption{Simulational results in 2 dimensions. The initial ball has 321 pixels
and parameters are $E = 2000$, $r = 80$ and $k = 0.2$ (see text). All fragments
obtained from 5000 independent falls were superimposed. The same behaviour is
obtained, with slightly different slopes, for many other combinations of these
parameters.}

\label{fig2}
\end{figure}

	In short, we have measured a crossover between two different power-laws
from large to short fragments, within a mercury ball crashing on the floor,
figure 1. We raise here the proposal that it is a real phenomenon, not a simple
undercounting artifact. We also introduced a dynamic stochastic model to
simulate this phenomenon in computers, which reproduces again the same
crossover in both 2 and 3 dimensions, figures 2 and 3.

	We would like to acknowledge CNPq, CAPES, FAPERJ and the
PRONEX-CNPq-FAPERJ program, under contract 171.168-2003, for partial financial
support.

\begin{figure}

\begin{center}
 \includegraphics[angle=-90,scale=0.5]{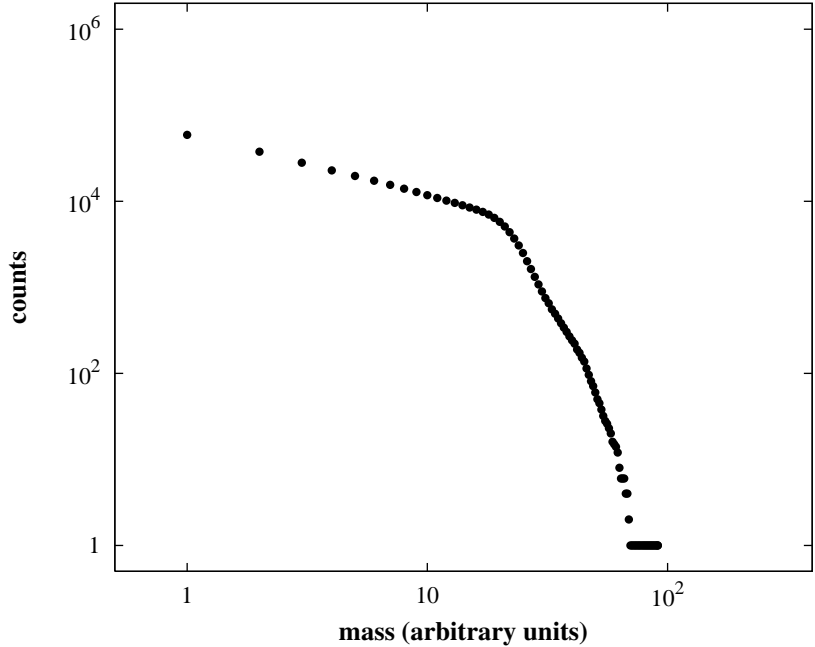}
\end{center}

\caption{Simulational results in 3 dimensions. The initial ball has 179
pixels and parameters are $E = 1$, $r = 50$ and $k = 0.9$, and $10000$
independent falls were superimposed. Again, the same behaviour with slightly
different slopes is obtained for many other combinations of these
parameters.}

\label{fig3}
\end{figure}


\newpage
\begin{thebibliography} {99}

\bibitem{Oddershede} L.B. Oddershede, PhD thesis, Odense University (1998);
L.B. Oddershede, P. Dimon and J. Bohr, {\it Phys. Rev. Lett.} {\bf 71}, 3107
(1993).

\bibitem{Astrom} J.A. {\AA}str\"om, F. Ouchterlony, R.P. Linna and J. Timonen,
{\it Phys. Rev. Lett.} {\bf 92}, 245506-1 (2004).

\bibitem{Campi} X. Campi, {\it J. Phys.} {\bf A19}, L917 (1986); M.L. Gilkes
{\sl et al}, {\it Phys. Rev. Lett.} {\bf 73}, 1590 (1994); J.B. Elliot {\sl et
al}, {\it Phys. Rev.} {\bf C55}, 1319 (1997); L. Beaulieu {\sl et al}, {\it
Phys. Rev. Lett.} {\bf 84}, 5971 (2000); S. Das Gupta, A. Majumder, S. Pratt
and A. Mekjian, {\sl pr\'e-print nucl-th}/9903007; C.E. Aguiar, R. Donangelo
and S.R. Souza, {\it Phys. Rev.} {\bf C}, to appear (2006); F.P.M. dos 
Santos, R. Donangelo and S.R. Souza, {\sl pr\'eprint} (2006).

\bibitem{Kuhle} A. K\"uhle, A.H. S{\char'34}rensen, L.B. Oddershede, H. Busch,
L.T. Hansen and J. Bohr, {\it Z. Phys.} {\bf D40}, 523 (1997);

\bibitem{Oddershede2} L.B. Oddershede and J. Bohr {\it Physica Scripta} {\bf
T67}, 73 (1996).

\bibitem{Korsnes} R. Korsnes, S.R. Souza, R. Donangelo, A. Hansen, M. 
Paczuski and K. Sneppen, {\it Physica} {\bf A331}, 291 (2004).

\bibitem{Oddershede3} L.B. Oddershede, A. Meibom and J. Bohr {\it Europhys. 
Lett.} {\bf 43}, 598 (1998).

\bibitem{Sotolongo} O. Sotolongo-Costa, Y. Moreno-Vega J.J. Lloveras-Gonz\'ales 
and J.C. Antoranz, {\it Phys. Rev. Lett.} {\bf 76}, 42 (1996).

\bibitem{drop} P.M.C. de Oliveira, T.J.P. Penna, A.R. Lima, J.S. S\'a Martins,
C. Moukarzel and C.A.F. Leite, {\it Trends in Statistical Physics} {\bf 3}, 137
(2000).

\bibitem{Manna} S.S. Manna, H.J. Herrmann and D.P. Landau, {\it J. Stat. Phys.} 
{\bf 66}, 1155 (1992).

\bibitem{faucet} P.M.C. de Oliveira and T.J.P. Penna, {\it J. Stat. Phys.} {\bf
73}, 789 (1993); P.M.C. de Oliveira and T.J.P. Penna, {\it Int. J. Mod. Phys.} 
{\bf C5}, 997 (1994); T.J.P. Penna, P.M.C. de Oliveira, J.C. Sartorelli, W.M. 
Gon\c calves and R.D. Pinto, {\it Phys. Rev.} {\bf E52}, R2168 (1995); A.R. 
Lima, T.J.P. Penna and P.M.C. de Oliveira, {\it Int. J. Mod. Phys.} {\bf C8},
1073 (1997).

\bibitem{nuclear} P.M.C. de Oliveira, J.S. S\'a Martins and A.S. de Toledo {\it
Phys. Rev.} {\bf C55}, 3174 (1997); J.S. S\'a Martins and P.M.C. de Oliveira,
{\it Int. J. Mod. Phys.} {\bf C9}, 867 (1998); J.S. S\'a Martins and P.M.C. de
Oliveira, {\it Nucl. Phys.} {\bf A643}, 433 (1998).

\bibitem{hysteresis} P.M.C. de Oliveira and A.P. Guimar\~aes, {\sl
International Conference on Magnetism}, Cairns, Australia (1997).

\bibitem{multilayers} A.R. Lima, M.S. Ferreira, J. d'Albuquerque e Castro and 
R.B. Muniz, {\it J. Mag. Mag. Mat.} {\bf 226}, 666 (2001).



\end{thebibliography}
\end{document}